\documentclass[useAMS, usenatbib, usegraphicx]{mn2e}

\pubyear{2013}

\begin{document}

\title[Hemispherical power asymmetries in the WMAP7 low-resolution maps]
{Hemispherical power asymmetries in the WMAP 7-year low-resolution temperature and polarization maps}

\author[F. Paci {\it et al.}]
{F.~Paci$^{1}$\thanks{E-mail: fpaci@sissa.it}, A.~Gruppuso$^{2,3}$, F.~Finelli$^{2,3}$, A. De Rosa$^2$, N.~Mandolesi$^{2,4,5}$, and P.~Natoli$^{2,5,6}$ \\
$^1$ 
SISSA - Scuola Internazionale Superiore di Studi Avanzati, via Bonomea 265, I-34136 Trieste, Italy \\
$^2$ 
INAF-IASF Bologna, Istituto di Astrofisica Spaziale e Fisica Cosmica 
di Bologna, via Gobetti 101, I-40129 Bologna, Italy \\
$^3$
INFN, Sezione di Bologna, Via Irnerio 46, I-40126 Bologna, Italy \\
$^4$
Agenzia Spaziale Italiana, Viale di Villa Grazioli 23, I-00198 Roma, Italy \\
$^5$
Dipartimento di Fisica and Sezione INFN, Universit\`a degli Studi di Ferrara, via Saragat 1, I-44100 Ferrara, Italy \\
$^6$
Agenzia Spaziale Italiana Science Data Center, c/o ESRIN, via Galileo Galilei, Frascati, Italy
 }

\label{firstpage}

\maketitle

\begin{abstract}
We test the hemispherical power asymmetry of the WMAP 7-year low-resolution temperature and polarization maps.
We consider two natural estimators for such an asymmetry and exploit our implementation of an 
optimal angular power spectrum estimator for all the six CMB spectra. By scanning the whole sky through a sample
of 24 directions, we search for asymmetries in the power spectra of the two hemispheres,
comparing the results with Monte Carlo simulations drawn from the WMAP 7-year
best-fit model. Our analysis extends previous results to the polarization sector. 
The level of asymmetry on the ILC temperature map is found to  be
compatible with previous results, whereas no significant asymmetry on the polarized spectra is detected. We show that our results are only
weakly affected by the {\it a posteriori} choice of the maximum multipole considered for the analysis.
We also forecast the capability to detect dipole modulation by our methodology at Planck sensitivity.

\end{abstract}

\begin{keywords}
cosmic microwave background - cosmology: theory - methods: numerical - methods:
statistical - cosmology: observations
\end{keywords}

%%%%%%%%%%%%%%%%%%%%%%%%%%%%%%%%%%%%%%%%%%%%%%%%%%%%%%%%%%%%%%%%%%%%%%%%
%                                                                                          INTRODUCTION 
%%%%%%%%%%%%%%%%%%%%%%%%%%%%%%%%%%%%%%%%%%%%%%%%%%%%%%%%%%%%%%%%%%%%%%%%

\section{Introduction}

Since the first claim of hemispherical power asymmetry in the Cosmic Microwave Background (CMB) temperature anisotropies \citep{Eriksen:2003db} as measured by the WMAP satellite 
in its first-year observation, a growing number of papers on the subject have appeared in the literature. Further investigations 
have refined the analysis and extended it to WMAP 3- and 5-year data \citep{Hansen:2004vq,Eriksen:2007pc}, leading to confirmation of a 
hemispherical power asymmetry - defined by Galactic 
coordinates $(\theta=107^\circ, \phi=226^\circ)$ - in the multipole range $\ell= [2,600]$, whose significance is as high a $99.6\%$ \citep{Hansen:2008ym}. 

As an alternative to a discontinuous change of the power on two opposed hemispheres, a dipolar modulation has also been considered, in the literature. 
In \citet{Hoftuft:2009rq} the data resolution is lowered and a modulation of the CMB signal is assumed for angular scales up to a maximum multipole considered, $\ell_{\rm{max}}$. 
Instead, \citet{Hanson:2009fd} proposed the use of an optimal quadratic estimator on full-resolution data, 
finding mild evidence for a dipolar modulation at large angular scales. They show that the effect strongly depends on the choice of $\ell_{\rm{max}}$ and that it decreases if the higher multipoles are included in the analysis.
A re-analysis of the latter has been presented by the WMAP team \citep{Bennett:2010jb}, where the connection between the asymmetry and the cutoff scale is further investigated.

However, the polarization sector remains poorly explored in this context.
\citet{Paci:2010wp} investigated the properties of the CMB polarization field on the two hemispheres defined by $(\theta=107^\circ, \phi=226^\circ)$ \citep{Hansen:2008ym}, exploiting an implementation of the
quadratic maximum likelihood (QML) method \citep{Gruppuso2009}.
No significant anomalies in the polarization and temperature-polarization cross-correlation were found in WMAP 5-year data \citep{Paci:2010wp}.
Dipolar modulations in temperature and polarization have been studied in \cite{Dvorkin:2007jp}, and more recently in \cite{Ma:2011ii}. 

In the present work we test the hemispherical power asymmetry at large scale on WMAP 7-year temperature and polarization maps without any theoretical assumptions. 
By sampling the whole sky in 24 equally-spaced symmetry
axis, we test the power asymmetry  on as many pairs of hemispheres. For each of those, we compute the same figures of merit as we did in our previous work \citep{Paci:2010wp}.
Moreover, we analyze the dependence of our results from the parameter $\ell_{max}$ along the lines suggested in \citet{Bennett:2010jb}, showing that our 
conclusions are only mildly affected by any {\it a posteriori} choice.

This paper is organized as follows. In
Section 2 we describe the methodology, the estimators and the dataset of our analysis. In Section 3 we present our
results assessing their significance by Monte Carlo simulations. In Section 4 we discuss the implications of power asymmetries in polarization for non-isotropic models.
In Section 5 we draw our main conclusions.

%%%%%%%%%%%%%%%%%%%%%%%%%%%%%%%%%%%%%%%%%%%%%%%%%%%%%%%%%%%%%%%%%%%%%%%
%                                                                         DESCRIPTION OF THE ANALYSIS                                        
%%%%%%%%%%%%%%%%%%%%%%%%%%%%%%%%%%%%%%%%%%%%%%%%%%%%%%%%%%%%%%%%%%%%%%%

\section{Description of the analysis}

In this section we review the algebra of the QML estimator, we define the 24 pairs of hemispheres under investigation and describe our dataset, simulations and estimators.

\subsection{Angular Power Spectra Estimation}

In order to evaluate the angular power spectra, we use the QML estimator.
The QML formalism was introduced in \citet{tegmark_tt} and extended to polarization in \citet{tegmark_pol}. 
In this section we describe the essence of the method. Further details
can be found in \citet{Gruppuso2009} where the {\it BolPol} code, our implementation of the QML estimator,  
has been applied to WMAP 5-year low-resolution data.

Given a map in temperature and polarization ${\bf x=(T,Q,U)}$, the QML provides estimates
$\hat {C}_\ell^X$ - with $X$ being one of $TT$, $EE$, $TE$, $BB$,
$TB$, $EB$ - of the angular power spectrum as: 

\begin{equation}
\hat{C}_\ell^X = \sum_{\ell' \,, X'} (F^{-1})^{X \, X'}_{\ell\ell'} \left[ {\bf x}^t
{\bf E}^{\ell'}_{X'} {\bf x}-tr({\bf N}{\bf
E}^{\ell'}_{X'}) \right] \, ,
\end{equation}

where the Fisher matrix $F_{X X'}^{\ell \ell '}$ is defined as

\begin{equation}
\label{eq:fisher}
F^{\ell\ell'}_{X X'}=\frac{1}{2}tr\Big[{\bf C}^{-1}\frac{\partial
{\bf C}}{\partial
  C_\ell^X}{\bf C}^{-1}\frac{\partial {\bf C}}{\partial
C_{\ell'}^{X'}}\Big] \,,
\end{equation}

and the ${\bf E}^{\ell}_X$ matrix is given by

\begin{equation}
\label{eq:Elle}
{\bf E}^\ell_X=\frac{1}{2}{\bf C}^{-1}\frac{\partial {\bf C}}{\partial
  C_\ell^X}{\bf C}^{-1} \, ,
\end{equation}

with ${\bf C} ={\bf S}(C_{\ell}^X)+{\bf N}$ being the global (signal plus noise) covariance matrix and $C_{\ell}^X$ the fiducial power spectrum. 

Although an initial assumption for a fiducial power spectrum
$C_\ell^X$ is needed, the QML method provides unbiased estimates
of the power spectrum contained in the map regardless of the initial guess,
\begin{equation}
\langle \hat{C}_\ell^X\rangle=\bar{C}_\ell^{X} \,,
\label{unbiased}
\end{equation}
where the average is taken over the ensemble of realizations and $\bar{C}_\ell^{X}$ denotes the underlying model.
The covariance matrix associated to the estimates is the inverse Fisher matrix,
\begin{equation}
\langle \Delta\hat{C}_\ell^X \Delta\hat{C}_{\ell'}^{X'} \rangle= ( F^{-1})^{X \, X'}_{\ell\ell'} \,,
\label{minimum}
\end{equation}
and it does depend on the assumption for the fiducial power spectrum $C_\ell^X$: 
the closer the guess to the true power spectrum is, the closer are the error bars to minimum variance. 
According to the Cramer-Rao inequality, Eq. (\ref{minimum}) tells us that 
the QML has the smallest error bars. 
We thus call the QML an `optimal' estimator.

\subsection{Direction sampling}

\begin{figure}
\begin{center}
\includegraphics[width=0.82\columnwidth, angle=90]{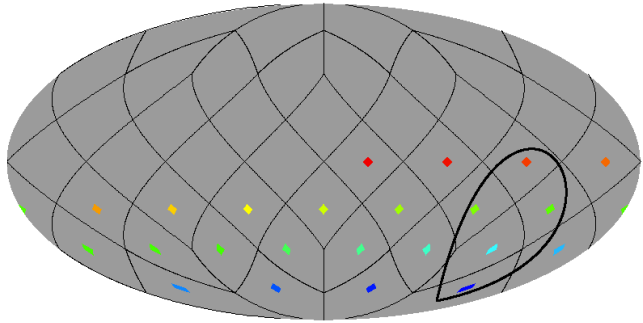}
\caption{Mollweide projection of the 24 directions defined by a Healpix $N_{\rm{side}}=2$ grid. The color scale runs from blue ($\hat{n}_{1}$) to red ($\hat{n}_{24}$). The black line defines the area of interest (see subsection \ref{subsec:angular}).}
\label{dirmap}
\end{center}
\end{figure}

In order to uniformly sample the sky, we have chosen 24 directions, as shown in Fig. \ref{dirmap}, defined by
the centers of a Healpix\footnote{http://healpix.jpl.nasa.gov/} \citep{gorski} $N_{\rm side}=2$ grid. 
Direction $\hat{n}_i$ will identify the axis defined by the $i$-th point of the grid.

For each of those, we have built a pair (North/South) of hemispherical masks at 
resolution $N_{\rm side}=16$. 
The hemispherical masks have been combined with the WMAP low-resolution Galactic masks for temperature (KQ85) and polarization (P06).

\subsection{Dataset and Simulations}

We use the temperature ILC map smoothed at $9.1285$ degrees and reconstructed at HealPix resolution $N_{\rm side}=16$, to which 
we add a random noise realization with variance of $1 \mu K^2$, as suggested in \citet{dunkley_wmap5}.
Consistently, the noise covariance matrix for $TT$ is taken to be diagonal with variance equal to $1 \mu K^2$.
For the polarization sector, we have adopted the $(Q,U)$ foreground cleaned low-resolution dataset publicly available at the LAMBDA website\footnote{http://lambda.gsfc.nasa.gov/}. 
Frequency maps ($m_i$) and covariance matrices ($C_i$) have been co-added as follows,
\begin{equation}
C_{tot}^{-1}=\sum_iC_i^{-1} \,, \qquad m_{tot}=C_{tot}\sum_i C_i^{-1}m_i
\end{equation}
where $i=Ka$, $Q$ and $V$.
Maps and covariances for the two sky regions (namely North and South) have been consistently tailored to the combined masks.

To assess the significance of the power asymmetries found in the data, our results have been tested against Monte Carlo simulations. 
A set of 10000 CMB+noise sky realizations has been generated: the signal was generated from the WMAP 7-year best-fit model \citep{Komatsu:2010fb}, the noise through 
a Cholesky decomposition of the global ($T,Q,U$) noise
covariance matrix. We have then computed the angular power spectra for each of the 10000 simulations using {\it BolPol} and built two figures of merit as explained in the next subsection.

\subsection{Estimators}

We define the following quantities
\begin{equation}
C^{X}_{N/S} \equiv {1\over {(\ell_{\rm{max}}-1)}} \, \sum_{\ell=2,\ell_{\rm{max}}} {\ell (\ell + 1) \over{2 \pi }} \, \hat{C}^{X, N/S}_{\ell} 
\label{CNS}
\end{equation}
where $\hat{C}^{X, N}_{\ell}$ and $\hat{C}^{X, S}_{\ell}$ are the estimated angular power spectra obtained with {\it BolPol} observing only the Northern (`$N$') and the Southern (`$S$') hemisphere respectively, outside the galactic plane. 
As above, $X$ runs over the spectral types. 

Two estimators can be built as follows: 
the ratio $R^X$, as performed in \citet{Eriksen:2003db},

\begin{figure}
\begin{center}
\includegraphics[width=\columnwidth]{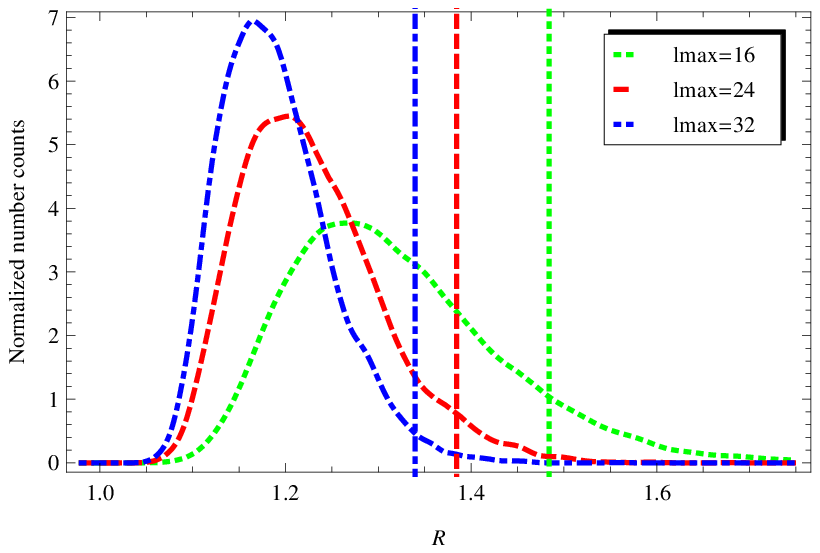}\\
\includegraphics[width=\columnwidth]{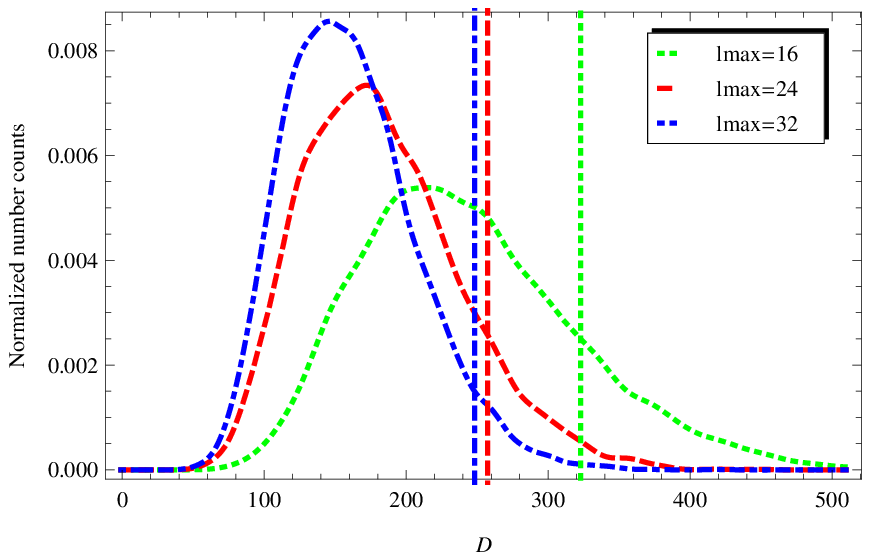}
\caption{Top panel: estimator $R$ for $TT$ computed for $\ell_{\rm{max}}=16$ (dotted green), 24 (dashed red), 32 (dot-dashed blue), for direction $\hat{n}_{14}$, the one closer to the axis found by \citet{Hansen:2008ym}.
The vertical lines show the value we extract from the WMAP data whereas the smoothed probability distribution are drawn from Monte Carlo simulations. Bottom panel: same for estimator $D$.}
\label{plotdir14}
\end{center}
\end{figure}

\begin{equation}
R^X = \textrm{max} \{C^X_S/C^X_N \, , C^X_N/C^X_S\}
\label{rapporto}
\end{equation}
and the difference $D^X$,
\begin{equation}
D^X=| C^X_S - C^X_N |\, ,
\label{differenza}
\end{equation}
of the two aforementioned quantities. 
In the following, we will drop the index $X$ for $R$ and $D$, and mention explicitly the spectrum we refer to.

For our application to WMAP data, both estimators may be considered for $TT$, while only the $D$ estimator 
has been applied to the other spectra ($EE$, $TE$, $BB$, $TB$ and $EB$), because of unfavorable 
signal-to-noise ratio of the WMAP data in polarization.

%%%%%%%%%%%%%%%%%%%%%%%%%%%%%%%%%%%%%%%%%%%%%%%%%%%%%%%%%%%%%%%%%%%%%%%
%                                                                                          RESULTS  
%%%%%%%%%%%%%%%%%%%%%%%%%%%%%%%%%%%%%%%%%%%%%%%%%%%%%%%%%%%%%%%%%%%%%%%

\section{Results}

\begin{table}
\caption{Probability for $R$ and $D$ of having a smaller value with respect to the WMAP one along direction $\hat{n}_{14}$.}
\centering 
\begin{tabular}{c c c c} 
\hline\hline 
 $TT(\hat{n}_{14})$ &  $\ell = 2-16$ & $\ell = 2-24$ & $\ell = 2-32$ \\ 
\hline 
$R$ & 90.94 & 96.16 & 98.38 \\
$D$ & 84.63 & 89.86 & 95.57 \\
\hline \hline
\end{tabular}
\label{tablevaluesdir14} 
\end{table}

\begin{figure}
\begin{center}
\includegraphics[width=\columnwidth]{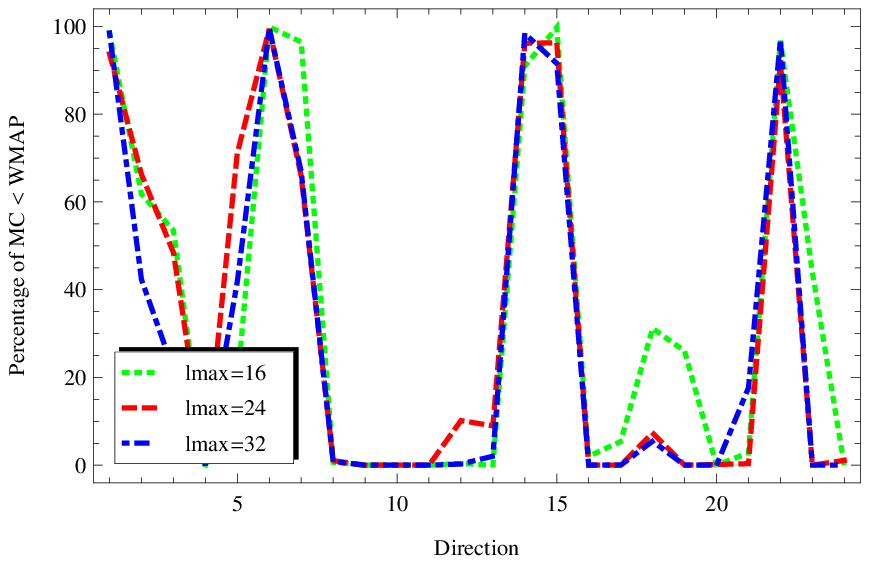}\\
\includegraphics[width=\columnwidth]{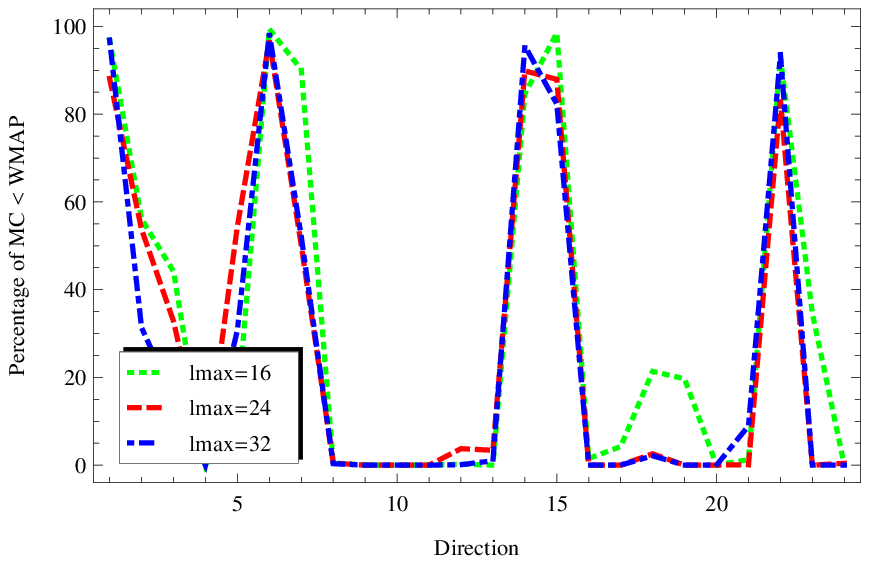}
\caption{Top panel: percentage asymmetry of the estimator $R$ for $TT$ computed for $\ell_{\rm{max}}=16$ (dotted green), 24 (dashed red), 32 (dot-dashed blue) along each of the 24 directions considered. Bottom panel: same for estimator $D$.}
\label{plotRDdirTT}
\end{center}
\end{figure}

As preliminar result, we report in Fig. \ref{plotdir14} our estimate of the temperature hemispherical asymmetry defined by direction $\hat{n}_{14}$ (the one within our sample which lies closer to the axis found by \citet{Hansen:2008ym}), for three values of the maximum multipole considered, taken as illustrative, $\ell_{\rm{max}}=16,$ 24, 36.
The two panels show the value of $R$ and $D$ as computed for the WMAP maps, compared to the probability distributions we have drawn from Monte Carlo simulations. The corresponding level of asymmetry is explicitly shown in Table \ref{tablevaluesdir14} for the corresponding multipole intervals. The Monte Carlo distribution does not depend on the specific direction considered, as the estimators are computed by maximizing the asymmetry over the 24 directions under investigation (see Subsection 3.2 for more details).

\begin{figure}
\begin{center}
\includegraphics[width=\columnwidth]{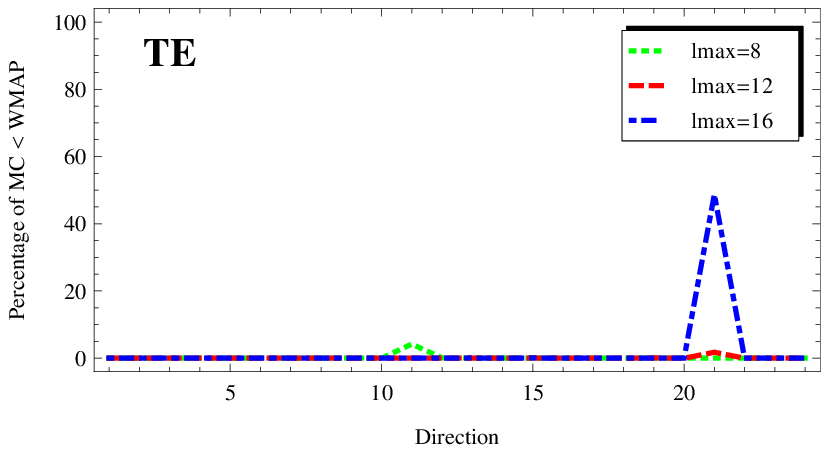}\\
\includegraphics[width=\columnwidth]{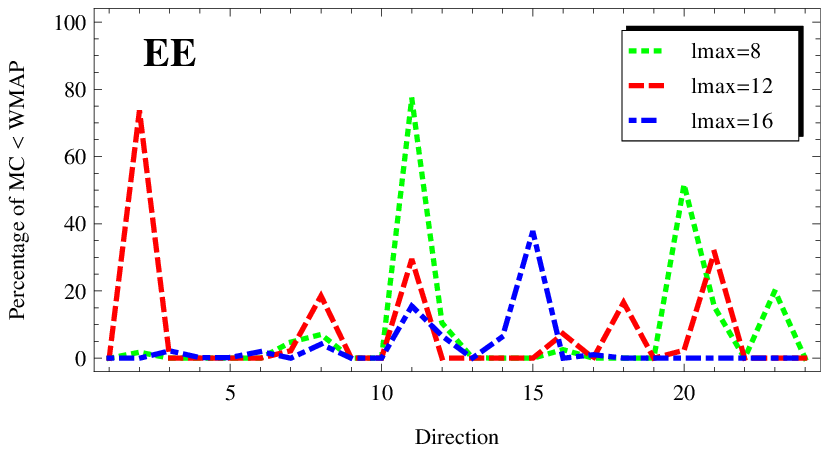}\\
\includegraphics[width=\columnwidth]{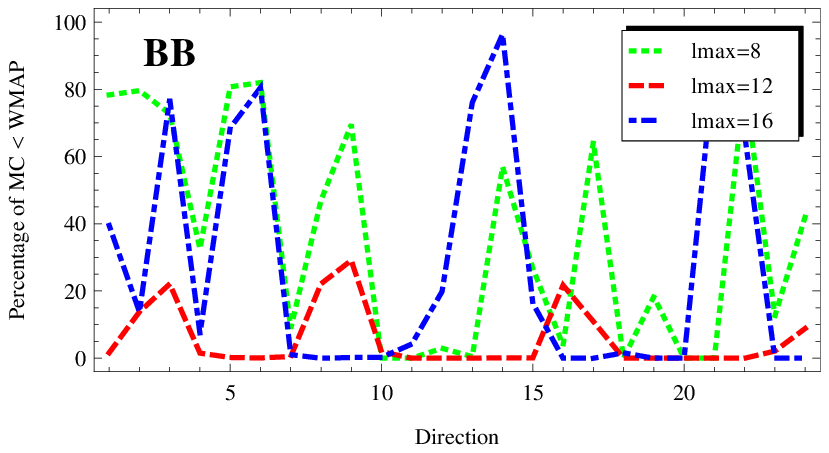}\\
\includegraphics[width=\columnwidth]{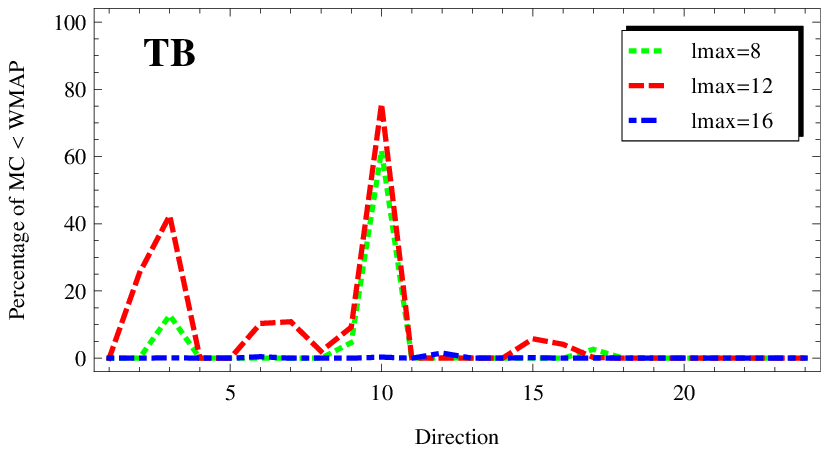}\\
\includegraphics[width=\columnwidth]{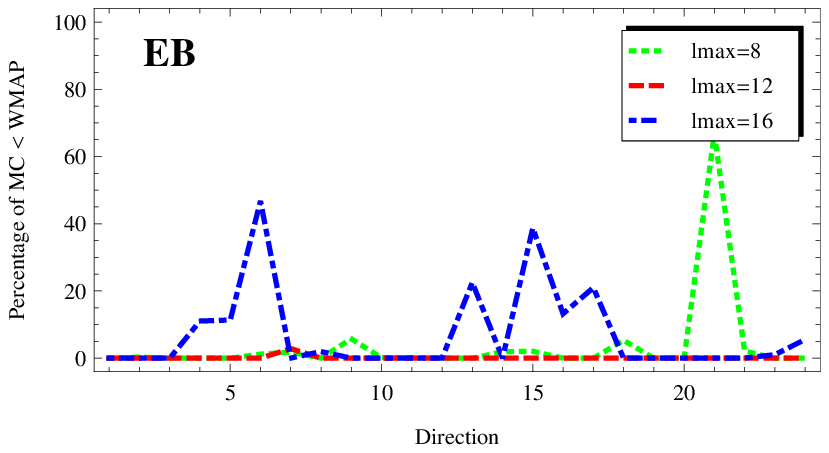}
\caption{Percentage asymmetry of the estimator $D$ computed for $\ell_{\rm{max}}=8$ (dotted green), 12 (dashed red), 16 (dot-dashed blue) along each of the 24 directions considered. The five panels are for $TE$, $EE$, $BB$, $TB$ and $EB$ respectively.}
\label{plotDdirPol}
\end{center}
\end{figure}

However, our extension of the same analysis to other direction in the sky suggests that also directions $\hat{n}_1$, $\hat{n}_6$, $\hat{n}_{15}$ and $\hat{n}_{22}$ have a comparable, or even higher, level of power asymmetry. This is shown in Fig. \ref{plotRDdirTT}, where we plot the percentage level of hemispherical asymmetry as defined by our sample of 24 directions for the same three maximum angular scales, $\ell_{\rm{max}}=16$, 24, 36. Those correspond to the region highlighted by the black circle in Fig. \ref{dirmap}. We will further investigate such region in the next subsection.

No significant hemispherical asymmetry is manifest in the cross- and polarization spectra (see Fig. \ref{plotDdirPol}), where the low signal-to-noise ratio reflects onto $\ell$-dependent fluctuations of $D$. We have also checked that $TB$ and $EB$ cross-spectra are well consistent with no asymmetry.

\begin{figure}
\begin{center}
\includegraphics[width=\columnwidth]{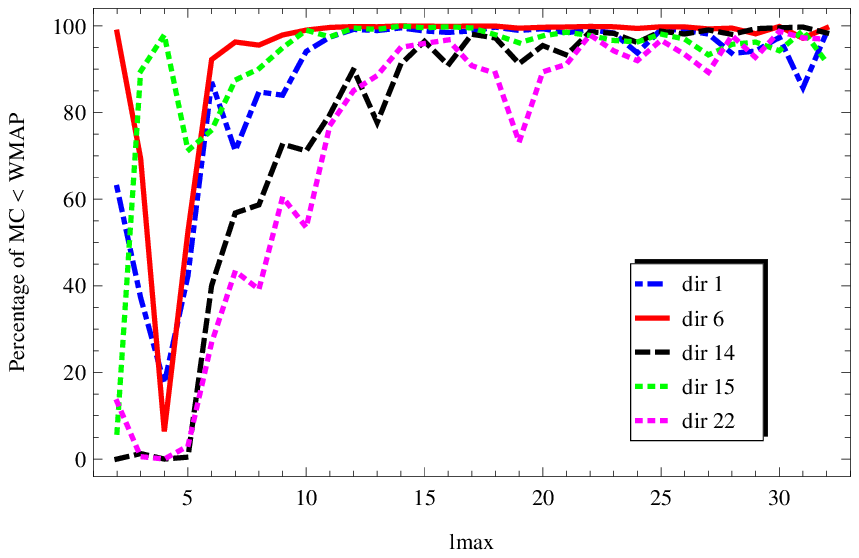}
\includegraphics[width=\columnwidth]{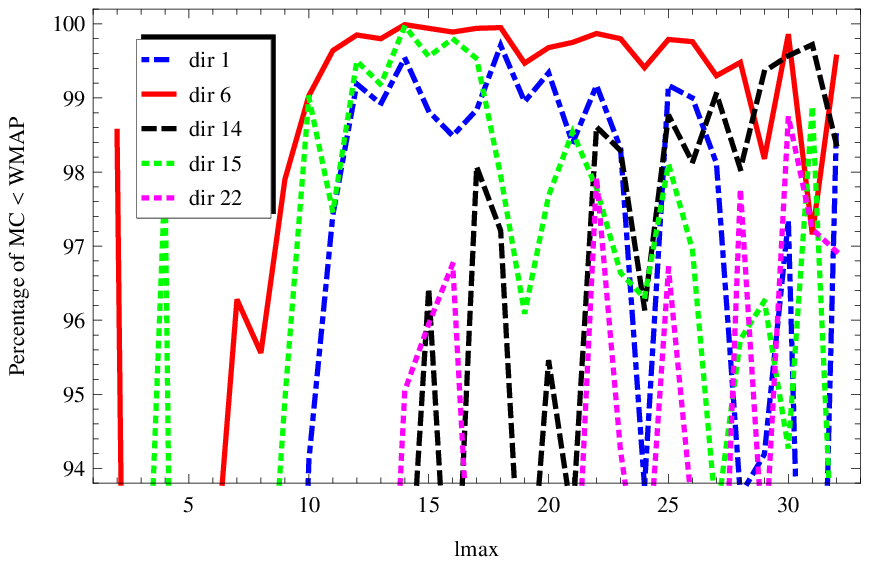} 
\caption{Percentage asymmetry of the estimator $R$ versus $\ell_{\rm{max}}$.  The most significative directions are shown:  $\hat{n}_1$ (dot-dashed blue), $\hat{n}_6$ (solid red), $\hat{n}_{14}$ (dashed black), $\hat{n}_{15}$ (dotted green), $\hat{n}_{22}$ (dotted magenta). Bottom panel shows a zoom of the top one.}
\label{RDellmax1}
\end{center}
\end{figure}

\begin{figure}
\begin{center}
\includegraphics[width=\columnwidth]{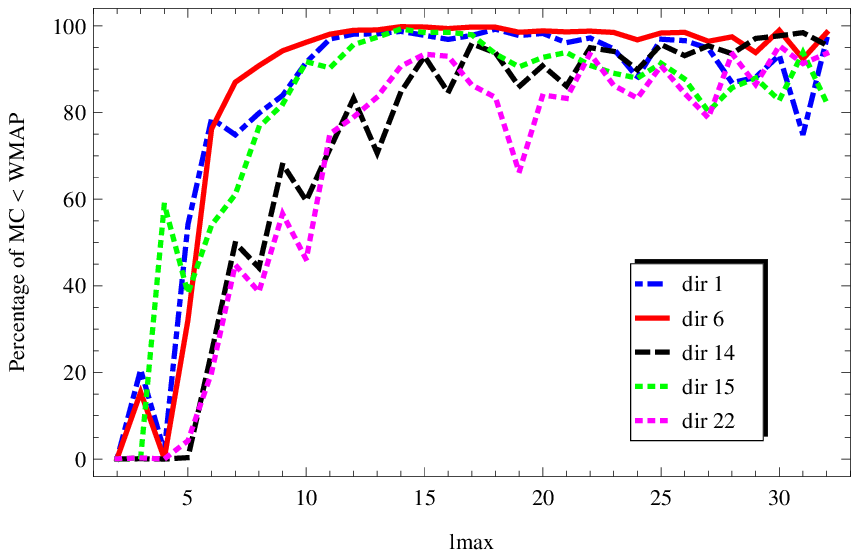}
\includegraphics[width=\columnwidth]{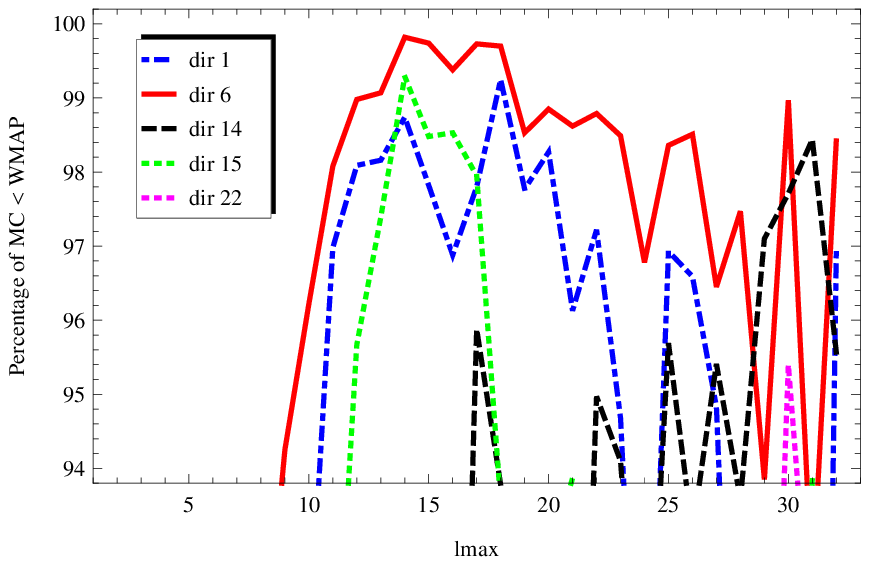}
\caption{Same as Fig. \ref{RDellmax1} but for $D$.}
\label{RDellmax2}
\end{center}
\end{figure}

\subsection{Angular scale dependence}
\label{subsec:angular}
We focus in this section on the most anomalous directions for the temperature field and investigate their multipole dependence. We restrict to directions $\hat{n}_1$, $\hat{n}_6$, $\hat{n}_{14}$, $\hat{n}_{15}$, $\hat{n}_{22}$ and let $\ell_{\rm{max}}$ vary from 2 to 32. Results are reported in Figs. \ref{RDellmax1},\ref{RDellmax2}. We observe that the cumulative power up to multipoles $\ell \sim 8$ does not show significant asymmetry, for none of the 5 hemispherical pairs considered. The only exceptions are $\ell=2$ and $\ell_{\rm{max}}=4$ for $\hat{n}_6$ and $\hat{n}_{15}$ respectively. We also notice that direction $\hat{n}_6$ (red curve), defined by Galactic coordinates ($\theta=132^\circ$, $\phi=245^\circ$), shows a constant, very high level of asymmetry through almost all the multipoles explored here, reaching the maximum for $\ell_{\rm{max}}=14$ as high as $99.99\%$ ($99.82\%$) for $R$ ($D$).

\subsection{Monte Carlo simulations}

Throughout the present work, uncertainties are assessed by Monte Carlo simulations \footnote{We sample our distributions by 10000 MC simulations. This introduces a resolution scale of $\sim 0.01\%$ in our assessments and prevents us from properly exploring effects at more than $\sim 3.5\sigma$.}. For each simulated CMB sky, the maximum asymmetry for a given angular scale does not necessarily lie along the direction which  maximizes the asymmetry in the data. Therefore, in order to properly sample the probability distribution from simulated skies, one has to maximize the asymmetry with respect to all the possible orientations, 24 in our analysis (see \citet{Finelli:2011zs} for a similar analysis in the context of mirror symmetry).
This is referred to as the  `look elsewhere' effect, and it has been properly taken into account in this work. 

\begin{figure}
\begin{center}
\includegraphics[width=\columnwidth]{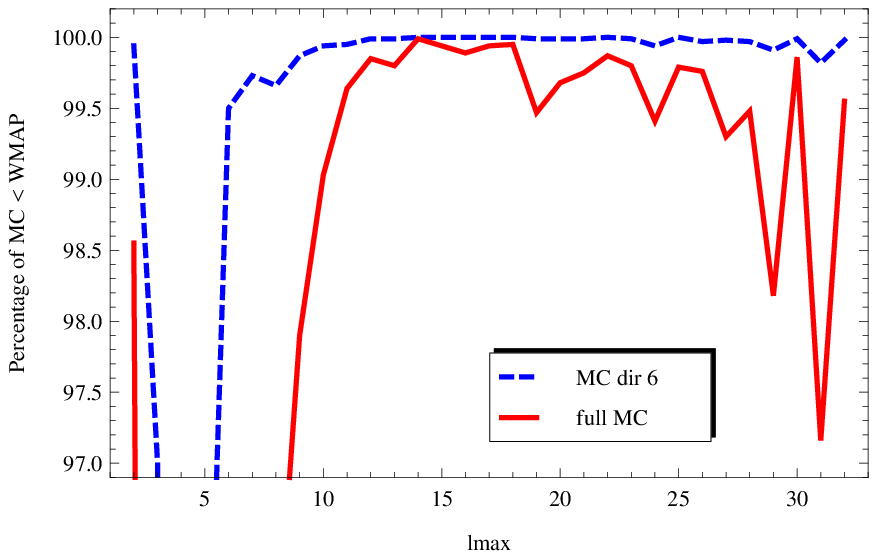}
\includegraphics[width=\columnwidth]{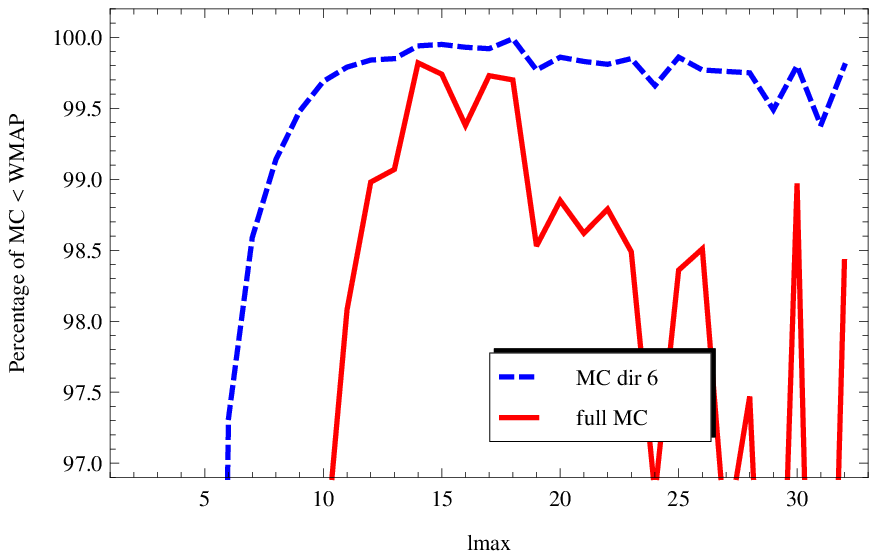}
\caption{Comparison of the percentage asymmetry we found on WMAP data along direction $\hat{n}_{6}$ (solid red curve) with what one would obtain if not accounting for the `elsewhere effect' (dashed blue curve). Top panel is for $R$, bottom panel for $D$.}
\label{MCtest}
\end{center}
\end{figure}

To stress the importance of this procedure, we show the impact on the estimated asymmetry of neglecting such effect. We fix one direction, $\hat{n}_{6}$, and compute $R$ and $D$ for 10000 CMB+noise simulated skies on the corresponding observed sky, without maximizing the estimators with respect to the other 23 directions. We compare then our distributions to WMAP asymmetry as estimated on the same sky fraction. 
The result of such a test is shown in Fig. \ref{MCtest}. The dashed blue curves show the asymmetry (as a function of $\ell_{\rm{max}}$) for the procedure just described, whereas the solid red curves refer to the correct analysis (see also Figs. \ref{RDellmax1},  \ref{RDellmax2}). As expected, neglecting the `look elsewhere effect' would lead to overestimate the significance of the asymmetry.

\subsection{Global statistical significance}
As pointed out by the WMAP team \citep{Bennett:2010jb}, care must be taken in assessing the significance of any claimed anomaly in the data, such as the hemispherical power asymmetry. 
We propose here the analogous of the analysis suggested in \citet{Bennett:2010jb} for the dipole power asymmetry. The idea is to associate a probability to our estimators as obtained from the WMAP 
dataset without any {\it a posteriori} choice of the $\ell_{\rm{max}}$ parameter. We want to compare the maximum probability ($\eta$) of asymmetry in the data to a distribution of probabilities 
drawn from MC simulations where $\ell_{\rm{max}}$ is chosen to maximize the asymmetry for each simulation independently. We let $\ell_{\rm{max}}$ vary from 2 to 32. We focus on the $D$ estimator for the temperature field. According to what shown in the previous subsection, we restrict our investigation to direction $\hat{n}_{6}$, for which the maximum value of $D$ is reached for $\ell_{\rm{max}}=14$ and corresponds to a significance $\eta^{WMAP}_D=99.82\%$.
Such a value has to be compared to the distribution of significance 
for the maximum asymmetry obtained from MC simulations ($\eta^{MC}_D$) where $\ell_{\rm{max}}$ is let free to move. In other words, for each extraction, $\ell_{\rm{max}}$ is chosen such that it maximizes the asymmetry of $D$.

\begin{figure}
\begin{center}
\includegraphics[width=\columnwidth]{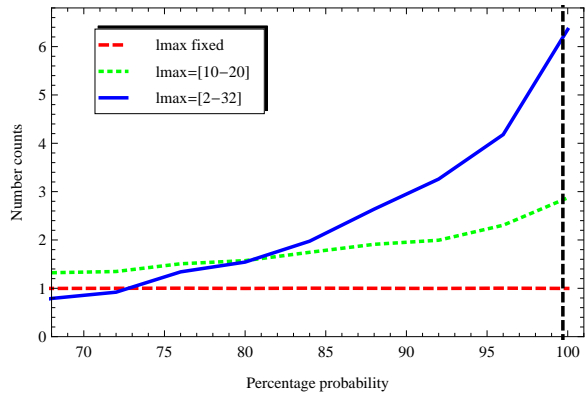}
\caption{Distribution of the significance for $\eta^{MC}_D$.
We plot the (normalized) number of simulations ($y$-axis) out of 10000 showing a given level of percentage asymmetry ($x$-axis) as computed by letting $\ell_{\rm{max}}$ vary for each simulation independently. 
Three cases are show: $\ell_{\rm{max}}=14$ (red), $10<\ell_{\rm{max}}<20$ (green) and $2<\ell_{\rm{max}}<32$ (blue). The black vertical line marks the WMAP value $\eta^{WMAP}_D=99.82\%$. }
\label{signif}
\end{center}
\end{figure}

Our results are reported in Fig. \ref{signif}, where the distribution of 10000 values of $\eta^{MC}_D$ is shown for three cases: $\ell_{\rm{max}}$ is kept fixed to a single (arbitrary) value (dashed red line), $12<\ell_{\rm{max}}<20$ (dotted green line) and $6<\ell_{\rm{max}}<32$ (solid blue line). The WMAP value is also reported as (dashed) black vertical line. As expected, we find that the probability for $\eta^{WMAP}_D$ to be anomalous decreases as we widen the range allowed for $\ell_{\rm{max}}$. However, even for $2<\ell_{\rm{max}}<32$, the probability that a MC simulation has a value of $\eta^{MC}_D$ that is larger than $\eta^{WMAP}_D$ is still as low as $1.76\%$.

\subsection{Implications of power asymmetry in polarization}
\label{subsec:dipolar}
We wish now to briefly discuss the implications of power
asymmetries in polarization at large angular scales for non-isotropic
cosmological models.
As a non-isotropic model, we consider the simple dipole modulation introduced in \cite{Gordon:2005ai} to explain 
the hemispherical asymmetry.

To assess the relative importance of polarization, we extend the model originally introduced for temperature to a full polarized 
($T,Q,U$) map by 

\begin{equation}
m_i = (1+ A \hat{p}_i \cdot \hat{n}) \bar{m}_{i} + n_i
\label{eq:dipole}
\end{equation}
where $\bar{m}_i$ is the ($T,Q,U$) isotropic contribution for the pixel $i$ pointing towards direction $\hat{p}_i$, $n_i$ is the instrumental noise contribution, $\hat{n}$ is the dipolar direction in the sky and $A$ is the amplitude of the dipolar effect. 

Whereas the WMAP instrumental noise prevents the information enclosed in the power
asymmetries in polarization to be a useful addition to the temperature
asymmetry generated by a dipolar modulation of Eq. (\ref{eq:dipole}), our estimator $D$ can be useful at Planck sensitivity.
In order to forecast the Planck capabilities, we follow Paci et al. (see also \cite{Ma:2011ii}) and assume uncorrelated uniform instrumental noise, whose amplitude is consistent with the Planck 143GHz channel sensitivity, $\sigma_T\sim0.2\mu$K and $\sigma_P\sim0.4\mu$K for temperature and polarization respectively \citep{bluebook}.
We choose one direction in the sky, i.e. $\hat{n}=\hat{n}_6$, and compute the difference power on the two hemispheres defined by the same direction for 10000 simulated skies. As amplitude we choose $A=0.114$, according to a previous analysis relying on a different estimator, but based on WMAP data at the same resolution and Gaussian smoothing scale as in the current work \citep{Eriksen:2007pc}. 

\begin{figure}
\begin{center}
\includegraphics[width=\columnwidth]{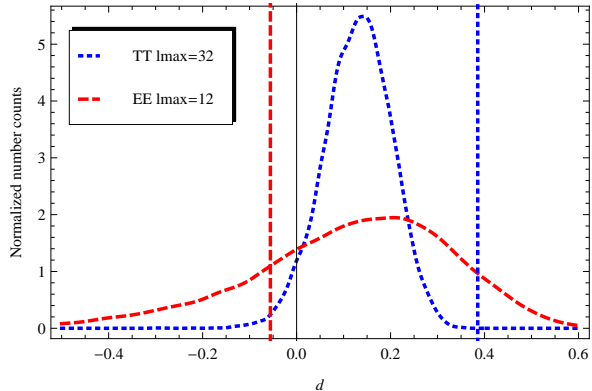}
\caption{Probability distribution of $d$ for $TT$ (blue dotted line) and $EE$ (red dashed line), for $\ell_{\rm{max}}$=32 and $\ell_{\rm{max}}$=12 respectively. Vertical lines shows the corresponding values as observed in the WMAP data.}
\label{dipole}
\end{center}
\end{figure}

Fig. \ref{dipole} shows the probability distribution of $d$, defined by
\begin{equation}
d^X= (C^X_S - C^X_N )/C^X_N \, ,
\end{equation}
for the $TT$ and $EE$ spectrum computed up to $\ell_{\rm{max}}=32$ and $\ell_{\rm{max}}=12$ respectively.
Our results demonstrate that polarization power spectra asymmetries will help in
characterizing better a dipolar modulation, even at the
$N_{\rm side}=16$ resolution with a temperature Gaussian smoothing of 9.1285 degrees: of course, due to the lower S/N with which CMB polarization 
is observed, polarization cannot compete with the capability to detect the dipole asymmetry by using temperature power asymmetry.
This relative importance of polarization with respect to temperature agrees with an analogous Planck forecast \cite{Ma:2011ii}, albeit for the 
different non-isotropic model based on a quadrupolar modulation.

It is also interesting to compare the current observational status with the predictions of the simple toy model in Eq. (\ref{eq:dipole}).
The WMAP 7-year $d$ estimators in temperature and $EE$ polarization displayed in Fig. \ref{dipole} have opposite sign, differently from  
the predictions of the simple toy model in Eq. (\ref{eq:dipole}) in which the asymmetries have the same sign (as also predicted by the
quadrupolar modulation in \cite{Ma:2011ii}). We also note that hemispherical power asymmetry in temperature $d$ predicted by 
Eq. (\ref{eq:dipole}) with $A=0.114$ - which corresponds to the results in \cite{Eriksen:2007pc} - 
is smaller than the one found on WMAP 7-year data by our QML analysis.

%%%%%%%%%%%%%%%%%%%%%%%%%%%%%%%%%%%%%%%%%%%%%%%%%%%%%%%%%%%%%%%%%%%%% 
%                                                                                          CONCLUSIONS 
%%%%%%%%%%%%%%%%%%%%%%%%%%%%%%%%%%%%%%%%%%%%%%%%%%%%%%%%%%%%%%%%%%%%%

\section{Conclusions}
We have performed for the first time a scan of the CMB sky in temperature and polarization searching for hemispherical asymmetries at low multipoles. 
We have sampled the whole sky in 24 directions and computed the six angular power spectra of the CMB through our implementation of the QML estimator. 
As adopted in previous analyses, we have estimated the hemispherical asymmetry along each axis as the ratio and the difference of the angular power up to a 
given scale $\ell_{\rm{max}}$, which we use as free parameter in the analysis. 
We find the maximum hemispherical asymmetry of the temperature field if the symmetry axis is chosen along Galactic coordinates ($\theta=132^\circ$, $\phi=245^\circ$) and on angular scales $2\le\ell\le 14$. The significance of such an asymmetry is $99.82\%$ when computed through the estimator $D$. 
In order to support our findings, we have tested the impact of the {\it a posteriori} choice of $\ell_{\rm{max}}$ in the analysis. 
We find that the anomalous asymmetry is milder once the $\ell_{\rm{max}}$ parameter is released, although still as high as $98.24\%$. 
In the polarization sector, we find no significant hemispherical asymmetry, neither along the direction studied in our previous work, nor with respect to any other symmetry 
axis considered here. We have also tested the possibility of detecting a dipolar modulation of the CMB by our methodology at Planck sensitivity. 
Our analysis shows that simple estimators as the hemispherical power asymmetries constructed by the QML will be a useful complement to temperature for Planck.

%%%%%%%%%%%%%%%%%%%%%%%%%%%%%%%%%%%%%%%%%%%%%%%%%%%%%%%%%%%%%%%%%%%%% 
%                                                                                          ACKNOWLEDGEMENTS  
%%%%%%%%%%%%%%%%%%%%%%%%%%%%%%%%%%%%%%%%%%%%%%%%%%%%%%%%%%%%%%%%%%%%%

\section*{Acknowledgements}
We acknowledge the use of the BCX and SP6 at CINECA under the agreement INAF/CINECA and the use of computing facility at NERSC.
We acknowledge use of the HEALPix \citep{gorski} software and analysis package for
deriving the results in this paper.  
We acknowledge the use of the Legacy Archive for Microwave Background Data Analysis (LAMBDA). 
Support for LAMBDA is provided by the NASA Office of Space Science.
We acknowledge partial support by ASI through ASI/INAF Agreement I/072/09/0 for the {\sc Planck} LFI Activity of
Phase E2 and by MIUR through PRIN 2009.

%%%%%%%%%%%%%%%%%%%%%%%%%%%%%%%%%%%%%%%%%%%%%%%%%%%%%%%%%%%%%%%%%%%%% 
%                                                                                          BIBLIOGAPHY  
%%%%%%%%%%%%%%%%%%%%%%%%%%%%%%%%%%%%%%%%%%%%%%%%%%%%%%%%%%%%%%%%%%%%%

\end{document}